\documentclass[aoas,nameyear,noautosecdot,dvips]{arximspdf}
\usepackage{dcolumn,multirow}

\doi{10.1214/09-AOAS315}
\volume{4}
\issue{2}
\pubyear{2010}
\firstpage{849}
\lastpage{870}

\makeatletter
\newcolumntype{d}[1]{D{.}{.}{#1}}
\newcommand{\eqref}[1]{(\ref{#1})}
\renewcommand{\citep}[1]{(\citeauthor{#1} \citeyear{#1})}
\newcommand{\sefac}[2]{C_{#2}(#1)}
\newtheorem{corollary}{Corollary}[section]
\newtheorem{lemma}[corollary]{Lemma}
\newtheorem{proposition}[corollary]{Proposition}
\newproclaim{remarks}{Remarks}

\newcommand{\pcor}{\rho_{y \cdot w|z\mathbf{x}}}
\newcommand{\SE}{\operatorname{SE}}
\newcommand{\teeWtilde}{t_{\tilde{W}}}
\newcommand{\teeW}{t_{W}}
\newcommand{\blp}{\operatorname{Pj}}
\newcommand{\effW}{F_{W}}
\newcommand{\pcortilde}{\rho_{y \cdot \tilde{w}|z\mathbf{x}}}
\newcommand{\pcorzw}{\rho_{z \cdot w|\mathbf{x}}}
\makeatother

\begin{document}
\begin{frontmatter}

\title{The sensitivity of linear regression coefficients' confidence
limits to the omission of a confounder}
\runtitle{Sensitivity of linear regression coefficients}

\begin{aug}
\author[A]{\fnms{Carrie A.} \snm{Hosman}\thanksref{t0}\ead[label=e1]{chosman@umich.edu}},
\author[A]{\fnms{Ben B.} \snm{Hansen}\thanksref{t1}\corref{}\ead[label=e2]{ben.b.hansen@umich.edu}}
\and
\author[B]{\fnms{Paul W.} \snm{Holland}\ead[label=e3]{roberta.holland@att.net}}
\runauthor{C. A. Hosman, B. B. Hansen and P. W. Holland}
\affiliation{University of Michigan, University of Michigan\\ and Paul
Holland Consulting Corporation}
\address[A]{C. A. Hosman\\
B. B. Hansen\\
Department of Statistics\\
University of Michigan\\
Ann Arbor, Michigan 48109-1107\\
USA\\
\printead{e1}\\
\phantom{E-mail:\ }\printead*{e2}} %adresu isvedimo komanda gale!
\address[B]{P. W. Holland\\
Paul Holland Consulting Corporation\\
703 Sayre Drive\\
Princeton, New Jersey 08540 \\
USA\\
\printead{e3}}
\thankstext{t0}{Supported in part by NSF Grant SES-0753164.}
\thankstext{t1}{Supported in part by NSA Grant 07Y-177
and a Ford Minority Dissertation Fellowship.}
\end{aug}

% HISTORY:
\received{\smonth{4} \syear{2009}}
\revised{\smonth{10} \syear{2009}}

% ABSTRACT
%
\begin{abstract}
Omitted variable bias can affect treatment effect estimates obtained
from observational data due to the lack of random assignment to
treatment groups. Sensitivity analyses adjust these estimates to
quantify the impact of potential omitted variables. This paper presents
methods of sensitivity analysis to adjust interval estimates of
treatment effect---both the point estimate and standard
error---obtained using multiple linear regression. Central to our
approach is
what we term \textit{benchmarking}, the use of data to establish
reference points for speculation about omitted confounders. The method
adapts to treatment effects that may differ by subgroup, to scenarios
involving omission of multiple variables, and to combinations of
covariance adjustment with propensity score stratification. We
illustrate it using data from an influential study of health outcomes
of patients admitted to critical care.
\end{abstract}

% KEYWORDS
%

\end{frontmatter}
%
%s1 ###
\section{Introduction}
\label{sec:introduction}
%s1.1 ###
\subsection{\texorpdfstring{Methodological context.}{Methodological context}}
In a common use of multiple linear regression, one regresses an
outcome variable on a treatment variable and adjustment variables,
then interprets the fitted treatment-variable coefficient as an
estimate of the treatment's effect on the outcome. The interpretation
relies on the
assumptions of the linear model and some assumption to the effect that
there either are no unmeasured confounders or at least none that
demand adjustment. (Ignorability of treatment assignment
[\citet{rubin1978}; \citet{holland1988a}, Appendix], is one such assumption;
there are many variants.)
The linearity assumptions are often testable given the data, but the
remaining assumption is not. When regression results are questioned,
it's often the nonconfounding assumption that is the focus of doubt.

Because the issue arises even with the most thorough observational
studies, adjusting for any number of covariates, it fuels cynicism
about observational research. If the possibility of unmeasured variable
bias can't be removed, then why bother with potential confounders,
particularly those that are difficult to measure, or not obvious
threats? It might be clear that the damage from omitting a confounder
$W$ would be reduced by adjustment for available correlates of $W,$
yet, because introducing these correlates would draw attention to the
absence of $W$, not at all clear that effecting the additional
adjustments would enhance the credibility of the research. Plainly, the
problem here is not the methodological strategy of broadly adjusting
for relevant baseline characteristics but an absence of, or lack of
awareness of, suitable methods with which to quantify benefits of more
comprehensive confounder controls.

Sensitivity analyses, procedures quantifying the degree of
omitted variable bias needed to nullify or reverse key conclusions of
a study, can help. Sensitivity analysis
methods for various models and data structures are proposed in
\mbox{\citet{cornfield1959}}, \citet{rr1983}, \citet{ros1988},
\mbox{\citet{copasli1997}}, \citet{robinsScharfRotn2000},
\citet{scharfstein2003gas}, \citet{marcus1997}, \citet{linetal98},
\citet{frank2000} and \citet{imbens2003sea}, among others, the last
four bearing closest resemblance to the approach to be presented here.
Invariably the methods start by in some way quantifying relationships
between hypothetical omitted variables and included ones, go on to
give an algorithm for converting these parameters into impacts on
estimates, $p$-values or confidence limits, and then leave to researchers
themselves the task of deciding what parameter values are plausible or
relevant. Here we develop a method following the first parts of this
general recipe, but then doing a bit more to help researchers calibrate
intuitions about speculation parameters.

The resulting technique
possesses a combination of advantages, making it both uniquely
practicable and conducive to certain insights. First, it applies to
inferences made with ordinary multiple regression, as we show in
Section \ref{sec:effect-estim-acco}, as well as to inferences made
with regression in combination with propensity-score stratification, a
topic discussed in Section \ref{sec:ppty-adjusted}. Second, it
quantifies relationships between omitted and included variables in terms
intrinsic to multiple regression, permitting intuitions for the
relationships to be calibrated with a few additional regression fits
(Section \ref{sec:effect-estim-acco}). [\citet{angristimbens2002}, Section
4,
suggest such calibration in a related context.] Third, it represents
effects of an omission with just two quantities, one tracking
confoundedness with the treatment variable and the other measuring
conditional association with the outcome
(Section \ref{sec:effect-estim-acco}). Fourth, there are important
practical benefits to sensitivity analysis based on both of these
quantities---``dual'' sensitivity analyses, in
\citeauthor{gastwirth1998das}'s terminology
(\citeyear{gastwirth1998das})---in that analysis based only on
confoundedness with the treatment variable is likely to overstate
sensitivity. Our method makes this plain (Section \ref{sec:sens-interv}),
as we will demonstrate with a case study to be introduced presently,
although some other methods may obscure it.
Fifth, it gives closed-form characterizations of how confidence
intervals, as opposed only to estimates or hypothesis tests, could be
changed by inclusion of the omitted confounder
(Section \ref{sec:theor-results-underl}). Sixth, the method readily adapts
to analyses in which several omissions are suspected, or where
interactions with the treatment variable are used to handle possible
effect heterogeneity (Section \ref{sec:extens-applic}). Seventh, the same application
brings to light certain practical advantages of the use of propensity
scores which to our knowledge have not previously been noted
(Section \ref{sec:ppty-adjusted}).

%s1.2 ###
\subsection{\texorpdfstring{A case study.}{A case study}}
\label{sec:case-study}
Our application is to Connors et al.'s
(\citeyear{connors1996}) highly influential, and controversial, study
of the critical-care procedure known alternatively as Swan--Ganz,
pulmonary artery or right heart catheterization (RHC). RHC is a
procedure to perform continuous measurements of blood pressure in the
heart and large blood vessels of the lungs. Introduced in 1970, it
became standard procedure without first being tested in clinical
studies, as might be expected today, and empirical assessments that
were subsequently conducted failed to uncover evidence that it
improved medical outcomes [e.g., \citet{gore1987}; \citet{zion1990}].
However, these studies were criticized for insufficient confounder
controls. Using a large sample, good measures and extensive
adjustments for background variables, \citet{connors1996} echoed the
disappointments of the earlier assessments and went further, finding
RHC to worsen, rather than improve, both mortality and the duration of
treatment in critical care. Each of these studies used nonrandomized
data and is in some degree vulnerable to omitted variable bias.
Although the results of subsequent randomized trials have been largely
consistent with this picture
[\citet{rhodes2002}; \citet{sandhan2003}; \citet{shah2004}; \citet{richard2003}; \citet{pacman2005}], the
procedure remains a staple of critical care, and the surrounding
debate continues. This paper examines how the omission of covariates from
\citeauthor{connors1996}'s data might skew point and interval
estimates of RHC's effect on length of stay, in the process shedding
light on the degree of protection from omitted variables afforded by
included ones.

%s1.3 ###
\subsection{\texorpdfstring{The SUPPORT data.}{The SUPPORT data}}
\label{sec:support-data}
Connors et al.'s (\citeyear{connors1996}) data come from the
Study to Understand Prognoses and Preferences for Outcomes and Risks of
Treatments (SUPPORT). The study collected data on the decision-making
and outcomes of seriously ill, hospitalized adult patients. Patients
included in the study had to meet certain entry criteria and a
predefined level of severity of illness. All 5735 \mbox{SUPPORT} patients who
were admitted to an ICU in the first 24 hours of study were analyzed
together, for reasons detailed in \citet{connors1996}. Data include
initial disease categories upon admission, physiological measurements,
and demographic information. Patients were coded as having had RHC if
it was performed within the first 24 hours of the study.

Length-of-stay is the difference between a patient's recorded dates of
entry and exit from the hospital. Here we study the omitted variable
sensitivity of Connors et al.'s finding that RHC increases costs, by
lengthening stays in the hospital. For cost analysis it is logical to
compare lengths of stay irrespective of whether those stays ended in
death, as Connors et al. do and as we do also. Since a medical
procedure could, in principle, shorten stays only by increasing
mortality, comparisons like this one speak directly to economic effects
but not to health effects of the procedure. (A study focused on health
outcomes, as opposed to resource utilization, would most naturally
begin by analyzing survival and continue with analyses of patient
experience, including duration of stay in the hospital, that address
the issue of censoring by death [e.g., \citet{rubin2006causal}].
Such analyses require methods other than ordinary multiple regression,
however, and sensitivity analysis for them is somewhat beyond the scope
of this paper.)

%To analyze the SUPPORT data, it is logical to first assess whether and
%how receiving treatment or having certain pretreatment characteristics
%affect the probability of survival. As OLS models are typically not
%used to estimate survival, a full treatment of the issue is beyond the
%scope of this paper; we note, however, that \citeauthor{connors1996}
%found that RHC did not improve the chance of patient survival. Despite
%this finding, RHC is still widely used. To investigate the continued
%use of RHC, we examine whether the use of the procedure reduces the
%expense of hospital resources as measured by the length of a patient's
%stay. Length-of-stay is the difference, in days, between a patient's
%recorded dates of entry and exit from the hospital.

%s1.4 ###
\subsection{\texorpdfstring{First-pass regression results.}{First-pass regression
results}}
\label{sec:first-pass-regr}

Length-of-stay is right-skewed. We log-transform it
before regressing it on RHC and covariates. There are approximately~50
covariates for
which regression adjustments might be considered; a~backward stepwise
selection procedure reduces this number to 19, and estimates the RHC
effect as $0.11$: taking the log transformation into account, RHC
seems to increase lengths of stay by about $100*(\exp(0.11) - 1) =
12\%$. To reflect variability added by variable selection
[\citet{faraway1992}], we ran a nonparametric bootstrap to determine
null quantiles for regression parameters' $t$-statistics; in this
case, however, bootstrap $t$-quantiles were similar to quantiles of
the appropriate $t$-distribution. (Code for these and other computations
discussed in the paper appears in a
supplement [\citet{hosmanetal2010supp}].) Either way, a 95\% confidence
interval for the RHC effect ranges from 0.06 to 0.16, encompassing
increases of 6\% up to 18\%.

% Table \ref{pointest} presents the point estimate of
% the coefficient of RHC, the treatment variable, from the reduced
% model, and confidence intervals for the effect of RHC on hospital stay
% using a student $t$ quantile and a bootstrapped $t$ quantile.

% %Table as of 2/1/09
% \begin{table}
% \begin{center}
% \caption{Effect of RHC on length of stay: point estimate and 95\%
%confidence intervals}
% \begin{tabular}{c|ccc}
% \hline
% & Point estimate & Student $t$ quantile CI & Bootstrap $t$ quantile
%CI \\
% \hline
% Effect of RHC on & 0.11 & (0.06, 0.16) & (0.06, 0.16) \\
% length of hospital stay&&\\
% \hline
% \end{tabular}
% \label{pointest}
% \end{center}
% \end{table}

% \subsection{Outline}
% Our paper continues as follows.
% % Section \ref{sec:data-prel-results} introduces the
% % data and uses multiple regression in the ordinary way to estimate
% % treatment effects.
% Section \ref{sec:effect-estim-acco} demonstrates
% that plausible impacts of omitted variables can be pinned down with a
% few straightforward additional analytic steps. The algebra underlying
% the method is presented in Section \ref{sec:theor-results-underl},
% while Section \ref{sec:extens-applic} returns to
% \citeauthor{connors1996}, expanding the sensitivity analysis of the
% ordinarily covariance-adjusted effect estimates and comparing them to
% estimates for which covariate adjustment is effected with the help of
% propensity scores; Section \ref{sec:summary-discussion} concludes.

%s2 ###
\section{Effect estimates accounting for omitted variables}
\label{sec:effect-estim-acco}
To understand how an omitted variable, $W$, could affect the
coefficient on the treatment variable, $Z$, in a hypothetical
regression of a given outcome on these and other variables,
%
%e1 ###
\begin{equation}\label{eq:desiredresult}
Y={\alpha}{\mathbf{X}}^T +{\beta}Z + {\delta}W + e  ,
\end{equation}
it is well to begin by examining how included variables affect the
treatment coefficient in the regression that was actually performed,
%
%e2 ###
\begin{equation}\label{eq:studyresult}
Y={a}{\mathbf{X}}^T+ {b}Z + e .
\end{equation}
This process lends context to an accompanying sensitivity analysis.

%s2.1 ###
\subsection{\texorpdfstring{Omitted variable bias.}{Omitted variable bias}}
\label{ssec:OVbias}
Perhaps the most familiar way of comparing regressors is in terms of
their effects on $R^{2}$; we begin in this mode. Of 33 regressors
(columns of the design matrix corresponding to the set of 19
covariates, some of which are categorical with multiple levels),
the one that moves $R^{2}$ the most is ``DNR status,'' an indicator of
whether there was a do-not-resuscitate order on file for the subject
on the day that he first qualified for inclusion in the study. Without
it, $R^{2}$ would be reduced from 0.141 to 0.112; in contrast, removal
of the next-most predictive regressor %"cat1CHF"
reduces $R^{2}$ only to 0.131. On this basis one might expect DNR
status to have a relatively large effect on inferences about the RHC
effect, and in a sense it does: omitting it from the regression
equation increases the RHC coefficient estimate from 0.112 to 0.143, 1.2
standard errors. For comparison, consider a regressor which
contributes more modestly to the outcome regression: omission of
``bleeding in the upper GI tract,'' for instance, reduces $R^{2}$ only
by 0.001, the~28th smallest of 33 such reductions, and removing it
moves the treatment coefficient only a few percent of an SE. Blurring
the distinction between parameters and estimates just a little, we
refer to this difference as bias, the bias that would have been
incurred by omitting the DNR variable. It appears that $R^{2}$ does
track omitted-variable bias.

There is a simple relationship between a covariate $W$'s contribution
to explaining the response, as measured by its impact on $R^{2}$, and
the bias incurred by its omission. According to
Proposition \ref{prop:univarOVbias}, this bias is the product of the
standard error
on $Z$, the treatment variable, as calculated without regression
adjustment for~$W$ (another term to be explained presently) and
$\pcor$, the partial correlation of $W$ and the response given $Z$ and
remaining covariates, $X$. In turn, $\pcor^{2}$ equals
$[(1-R^{2}_{\mathrm{no}\mbox{ } W})- (1-R^{2}_{\mathrm{with}\mbox{ }
W})]/(1-R^{2}_{\mathrm{no}\mbox{ }W})$, the proportionate reduction in
unexplained variance when $W$ is added as a regressor [\citet{christensen1996}, Chapter 6].
That is to say, the square of the bias due to omitting a covariate is
linear in the fraction by which that covariate would reduce
unexplained variance. Returning to our example, DNR status consumes
3.3\% of outcome variation that would otherwise be unexplained, which
is 16 interquartile ranges greater than the third quartile of $\pcor
^{2}$'s associated
with the 33 available regressors; the $\pcor^{2}$ for upper GI
bleeding, in contrast, is 0.1\%, just above the first quartile.
% HERE'S WHERE I'D SUGGEST YOU INSERT A PARAGRAPH BREAK AND EXPLICITLY
% INTRODUCE THE TOPIC OF THE NEW PARAGRAPH BEFORE CONTINUING WITH THE
%FOLLOWING

The degree to which an omitted variable can bias effect estimates
through the value
of $\pcor^{2}$ depends on the value of $R^{2}$ with which we begin. In
our example,
the value of $R^{2}$ is relatively small; when the $R^{2}$ value is bigger,
changes in $R^{2}$ have a more pronounced effect. Imagining another
regression with a higher baseline $R^{2}$ and thus less unexplained
outcome variation,
consider an unmeasured covariate that reduces $R^{2}$ by the same percentage
as the DNR variable in our regression. As a result of having less
unexplained variation to start, the proportionate reduction in unexplained
variation, $\pcor^{2}$, must be larger than that of DNR in our example.
Such a
variable would consequently have a larger effect on the estimate of
omitted variable
bias.

%t1 ###
\begin{table}[b]
\caption{Selected included covariates' relation to treatment and
response variables, given remaining included variables}\label{modspecs}
\begin{tabular*}{\textwidth}{@{\extracolsep{\fill}}ld{2.1}c@{}}
\hline
& \multicolumn{1}{c}{\multirow{3}{90pt}{\centering \textbf{Confounding with RHC} $\bolds{(|\teeW|)}$ \textbf{(rounded)}}}
& \multicolumn{1}{c@{}}{\multirow{3}{104pt}[1pt]{\centering \textbf{\% decrease in unexplained variation by adding}\vspace*{1pt} $\bolds{W}$ $\bolds{(100\rho^2_{y \cdot
w|z\mathbf{x}})}$}}\\[20pt]
\hline
Income$^\ast$ & 6.8 & 0.3 \\
Primary initial disease cat.$^\ast$ & 48.1 & 3.4 \\
Secondary initial disease cat.$^\ast$ & 20.2 & 0.8 \\[3pt]
\textit{Comorbidities illness}:\\
Renal disease & 2.1 & 0.2 \\
Upper GI bleed & 0.7 & 0.1 \\[3pt]
\textit{Day 1 measurements}:\\
APACHE score & 5.1 & 0.1 \\
White blood cell count & 0.5 & 0.0 \\
Heart rate & 2.5 & 0.0 \\
Temperature & 2.3 & 0.1 \\
PaO2$/$FIO2& 15.4 & 0.1 \\
Albumin & 2.3 & 0.7 \\
Hematocrit & 3.3 & 0.9 \\
Bilirubin & 2.2 & 0.1 \\
Sodium & 3.1 & 0.1 \\
PaCo2 & 6.8 & 0.2 \\
DNR & 6.5 & 3.3 \\
PH & 3.7 & 0.3 \\[3pt]
\textit{Admit diagnosis categories}:\\
Neurology & 5.9 & 0.2 \\
Hematology & 3.6 & 0.1 \\
\hline
\end{tabular*}
\tabnotetext{}{Categorical variables with
multiple levels. (Strictly speaking, these rows of the table give
${F_{W}^{1/2}}$ in the middle column, not $t_{W}$; see Section \protect\ref{sec:severalOVs}.)}
\end{table}

The remaining factor in Proposition  \ref{prop:univarOVbias}'s
expression for
omitted variable bias expresses the degree to which the variable is
confounded with treatment. This confounding turns out to figure more
centrally in omitted variable bias, despite the fact that measurements
of it arise less often in regression analysis than do $\pcor$ or
$R^{2}$. A straightforward descriptive measurement falls out of the
regression of $Z$ on the remaining regressors from the outcome
regression: their confoundedness with $Z$ is reflected in their
$t$-statistics, the ratios of their coefficient estimates to their
conventional standard errors. Note that these are here considered as
descriptive, not inferential, statistics; were this auxiliary
regression instead being used for statistical inference, a nonlinear
model may be required for binary $Z$, in which case these nominal
standard errors may not be appropriate. Denoting by $\teeW$ the
$t$-ratio attaching in this way to a regressor $W$, by $b$ the
RHC-coefficient in the absence of~$W$ and by $\beta$ the same
coefficient when $W$ is included, the bias due to omission of a
regressor decomposes as follows.
\begin{proposition}
\label{prop:univarOVbias}
If $R^2_{y \cdot z\mathbf{x}}<1$ and $\teeW$ is finite, then
%
%e3 ###
\begin{equation}
\label{eq:biasexpr}
\hat b - \hat\beta= \SE(\hat b)\teeW\pcor.
\end{equation}
\end{proposition}

A proof is given in Section \ref{sec:theor-results-underl}.

Besides being a strong predictor of the response, DNR status ranks
highly in terms of its confounding with the treatment: if RHC is
regressed on the covariates, then its $t$-statistic is $-6.5$, the fifth-largest
in magnitude. (Table \ref{modspecs} displays~$\teeW$ values, along with
a few square roots of $F$-statistics, to be explained in Section~\ref
{sec:severalOVs}, for multicategory variables. The magnitudes of the
$F$-statistics are not directly comparable to those of the $t$-statistics.)
Consistent with there being little bias associated with removing it,
upper GI bleeding
is as weakly related to the treatment variable as it was to the
response; its $t$ is
only $-0.7$, the fifth-smallest in magnitude. The strongest confounder
among the
adjustment variables is the PaO2$/$FIO2 ratio, a reflection of
oxygenation in the lungs, % cf.
%http://www.nurse.com/ce/course.html?CCID=3479 , "Acute lung injury"
%section.
with $t = -15.3$. On the other hand, its relatively small
contribution toward explaining variation in the outcome,
$\pcor^{2}=0.0007$, limits the effects of its removal on the treatment
coefficient; here $\hat b - \hat\beta$ is $0.011$, or 40\% of an SE.
Although the effect on the treatment coefficient of excluding the
PaO2$/$FIO2 variable is tempered by its small corresponding $\pcor^{2}$,
\eqref{eq:biasexpr} says that an actual omitted variable this
confounded with treatment could contribute up to 15.3 standard errors'
worth of bias (depending on how much more outcome variation it
explains). If this possibility is not merely notional, it follows
that uncertainty assessments based in the usual way on $\SE(\hat b) $
are quite incomplete.

Proposition \ref{prop:univarOVbias} suggests that the degree of an
omitted variable's confounding with the treatment plays a potentially
much larger role in determining the bias from its omission than does
its conditional association with the response. As it can explain no
more than 100\% of otherwise unexplained variation, in general
$\pcor^2 \leq1$, ensuring in turn that $|\hat b - \hat
\beta|/\SE(\hat b) \leq|\teeW|$. In contrast, nothing bounds the
$t$-ratios $\teeW$---covariates highly collinear with the treatment
can introduce biases that are arbitrarily large, at least in
principle. We refer to an unsigned statistic
$|\teeW|$ as the \textit{treatment confounding} of $W$ a covariate
or potential covariate $W$.

Table \ref{modspecs} reports $\teeW$ and $\pcor$ values that actually
occur in the SUPPORT data, placing observed covariates in the role of
$W$ one at a time. With these data, $\teeW$ is often of a magnitude
to make large biases possible (although this is somewhat tempered by
relatively small $\pcor$'s). The calculations permit the statistician
and his audience to calibrate intuitions for $\teeW$ against actual
variables and their relations to the treatment. We refer to this as
\textit{benchmarking} of treatment confounding.

%s2.2 ###
\subsection{\texorpdfstring{Omitted variables and the treatment's standard error.}{Omitted
variables and the treatment's standard error}}
\label{ssec:OVandSE}

The representation~(\ref{eq:biasexpr}) of omitted-variable bias, as the
product of $\SE(\hat b) $ with other factors, lends hope that a
meaningful combinations of errors due to variable omissions and
sampling variability might be bounded by a multiple of $\SE(\hat b) $.
A tempting solution is simply to add to the normal 97.5 percentile
1.96, or to whatever other multiplier of the SE has been found
appropriate to describe likely sampling error, a quantity large enough
to bound plausible values of $|\teeW\pcor|$. This solution will often
miss the mark, however, since omitting a covariate affects standard
errors as well as coefficient estimates. This is illustrated in
Table \ref{tab:3omits}, which shows effects on estimation of the
RHC-coefficient of omitting one-by-one the regressors discussed in Section
 \ref{ssec:OVbias}.

% latex table generated in R 2.7.1 by xtable 1.5-3 package
% Wed Feb 4 10:58:49 2009
%
%t2 ###
\begin{table}
\caption{RHC-coefficient and its standard error after removing
specified variables}
\label{tab:3omits}
\begin{tabular*}{265pt}{@{\extracolsep{\fill}}lccc@{}}
\hline
& &\multicolumn{2}{c@{}}{\textbf{Standard error}} \\[-6pt]
&& \multicolumn{2}{c@{}}{\hrulefill} \\
\textbf{Excluded  variable} & \textbf{Estimate} & $\bolds{\mathbf{df}=5700}$ &$\bolds{\mathrm{df} = 50}$\\
\hline
No exclusion & 0.112 & 0.0260& 0.278\\
DNR order & 0.143 & 0.0264& 0.206\\
GI bleed & 0.112 & 0.0260&0.274\\
PaO2$/$FIO2 & 0.122 & 0.0255&0.115 \\
\hline
\end{tabular*}
\end{table}

According to Table \ref{tab:3omits}, adjusting the RHC coefficient for
covariates including DNR order gives a smaller standard error, 0.0260,
than does excluding DNR order and adjusting for remaining covariates
only ($\operatorname{SE}=0.0264$). In contrast, inclusion of PaO2$/$FIO2 among the
adjustment variables has the effect of increasing the standard error,
from 0.0255 to 0.0260. Including or excluding upper GI tract bleeding
leaves the standard error unchanged, suggesting that omitted variables
associated with little bias should have little effect on the standard
error. This turns out to be true, in a limited sense, as can be read
from Proposition \ref{prop:univarOVSE}: the same statistics governing
the bias due to omission of a covariate $W$ also govern its effects on
the standard error, in such a way that when both $\teeW$ and $\pcor$
are small, then including or excluding $W$ has little effect on the
$Z$-coefficient's standard error.
\begin{proposition}
\label{prop:univarOVSE}
If $R^2_{y \cdot z\mathbf{x}}<1$ and $\teeW$ is finite, then
%
%e4 ###
\begin{equation}
\label{eq:univarOVSE}
\SE(\hat{\beta}) = \SE(\hat{b})\sqrt{1+ \frac{1 + \teeW^2}{\mathrm{df}-1}}\sqrt{1-\rho^2_{y \cdot w|z\mathbf{x}}}.
\end{equation}
Here $\mathrm{df}=n - \operatorname{rank}(X)-1$, the residual degrees of
freedom after $Y$ is regressed on $X$ and $Z$.
\end{proposition}

Proposition \ref{prop:univarOVSE} will be strengthened in
Proposition \ref{prop:mvOVSE}, which is proved in the \hyperref[app]{Appendix}.

Whereas
$|\teeW|$ and $|\pcor|$ both associate directly with the size of
the bias from omitting $W$, they act in opposite directions on the
standard error---reflecting the fact that increasing $R^2$ tends to
reduce coefficients' standard errors, except when it's increased at
the expense of introducing colinearity among regressors. The
difference explains why omitting some variables
increases standard errors, whereas omitting others decreases them.

It also reaffirms that omitted variables' effects on statistical
inferences are only incompletely reflected in omitted variable
bias. In the context of the present example, variable omissions have
only modest effects on the standard error; but this is a consequence
of the sample being quite large ($n=5735$) relative to the rank of the
covariate matrix (33): barring astronomically large $\teeW$s, with
$\mathrm{df}=5700$ the middle factor on the left side of \eqref{eq:univarOVSE}
has to fall close to 1; in consequence, even a $\teeW$ of 16 inflates
the standard error by at most 2\%, according to (\ref{eq:univarOVSE}).
In moderate and small samples, standard errors are more sensitive.
The third column of Table \ref{tab:3omits} presents standard errors as
they would be if the sample had been much smaller, such that $\mathrm{df}=50$,
but with the same sample means and covariances. While the omission of
a variable like GI bleed, which is only weakly related with
already included variables, still leaves the treatment's standard
error much the same, variables that are either moderately or strongly
confounded with the treatment now cause wide shifts in the magnitude
of the standard error. Adjustment for the PaO2$/$FIO2 measurement, for
example, now increases the treatment's standard error by a whopping~59\%.

%s2.3 ###
\subsection{\texorpdfstring{One sensitivity parameter or two?}{One sensitivity parameter or two?}}
Should we be inclined to consider worst-case scenarios,
Proposition \ref{prop:univarOVSE} reaffirms
Proposition \ref{prop:univarOVbias}'s message that the omitted
variable's treatment confounding, not its potential to increase
$R^{2}$, most demands our attention. The $\pcor$-contribution to
$\SE(\hat\beta)$ is a factor bounded above by 1, whereas the
$\teeW$-factor can be arbitrarily large---as could the
$\teeW$-contribution to omitted-variable bias. The greater potential
disturbance from confounding with the treatment than from large new
contributions to the outcome regression seems to be a feature multiple
regression analysis shares with other statistical techniques; indeed,
sensitivity analysis methods which set out to limit inference errors
in terms of a single sensitivity parameter parametrize confounding
with treatment, not predictivity of the response, countenancing
arbitrarily strong response-predictors in the role of the hypothetical omitted
variable [e.g., \citet{ros1988}; \citet{rosenbaumSilber2010}].

What would such an analysis suggest about the sensitivity of our RHC
results to the omission of a confounder not unlike included
variables?\break
PaO2$/$FIO2 was sufficiently confounded with treatment to indicate as
many as 16 standard errors' worth of bias. Yet its inclusion in the
regression adjustment moves the treatment effect by less than half a
standard error. To have moved $\hat\beta$ so much, it would have had
to consume all of the variation in $Y$ that was not explained without
it, something no one would have expected it to do. To the contrary,
inspection of other variables' contributions to $R^{2}$, as enabled by
Table \ref{modspecs}, suggests that $0.1$ (for instance) would be a
quite generous limit on $|\pcor|$. This in turn would restrict
omitted variable bias due to a treatment-confounder as strong as
PaO2$/$FIO2 to 1.6 SEs---still a meaningful addition to the
uncertainty estimate, if a less alarmist one. Rather than reducing
the number of sensitivity parameters by permitting $\pcor$ to fall
anywhere within its a priori limits, it is more appealing to retain
$\pcor$, restricting it within generous bounds of plausibility.
% Cite Rosenbaum & Silber Amplification paper here?

%s3 ###
\section{Sensitivity intervals}
\label{sec:sens-interv}
Taken together, Propositions \ref{prop:univarOVbias} and \ref
{prop:univarOVSE} enable a precise, closed-form description of the
union of interval estimates $\{ \hat\beta\pm q\SE(\hat\beta) \dvtx |\teeW
| \leq T, \pcor^2 \leq R \}$, for any nonnegative limits $T, R$ on
omitted variables' treatment confounding and contributions to reducing
unexplained variation---the collection of $Z$-slopes falling within
the confidence interval after addition of a covariate $W$ such that $-T
\leq\teeW\leq T$ and $\pcor^2 \leq R$.\vspace*{1pt} Such a union of intervals is
itself an interval: following \citet{ros2002}, we call it a \textit
{sensitivity interval}; and following \citet{small2007sai}, we refer to
the determining set of permissible values for $(\teeW, \pcor^2)$ as a
\textit{sensitivity zone}. The mapping of sensitivity zones to
sensitivity intervals is given in Proposition \ref{keyprop}.

\begin{proposition}
Let $Y$, $\mathbf{X}$, $Z$ and $W$ be as in \textup{(\ref{eq:desiredresult})}
and \textup{(\ref{eq:studyresult})}, with both regressions fit either by
ordinary least squares or by weighted least squares with common
weights. Assume $R^2_{y \cdot z\mathbf{x}}<1$. Let $\pcor$, $\teeW$
and $\mathrm{df}$ be as defined in Section \textup{\ref{sec:effect-estim-acco}}, fix $q>0$
and write $\sefac{t}{d}$ for $[1+(1+t^{2})/(d-1)]^{1/2}$.

\begin{enumerate}[(iii)]
\item[(i)] Assuming only $\teeW^{2} \leq T < \infty$,
%
%e6 ###
%e5 ###
\begin{eqnarray}
\hat{\beta} \pm q\operatorname{SE}(\hat{\beta}) &=& b \pm\bigl[-\teeW\pcor+ q
\sefac{\teeW}{\mathrm{df}}\sqrt{1-\pcor^{2}}  \bigr]\SE(\hat b) \label
{eq:climfmla}\\
&\subseteq&
b \pm\sqrt{T^{2} + q^{2}\sefac{T}{\mathrm{df}}^{2}}  \SE(\hat
b). \label{eq:si-norho}
\end{eqnarray}
\item[(ii)] Assuming $\teeW^{2} \leq T < \infty$, $\pcor^{2} \leq
R$ where $0 < R < T^{2}/(T^{2} + q^{2}\sefac{T}{\mathrm{df}}^{2})$,
%
%e7 ###
\begin{equation}
\hat{\beta} \pm q\SE(\hat{\beta})
\subseteq b \pm[TR^{1/2}+ q\sefac{T}{\mathrm{df}}(1-R)^{1/2}]\SE(\hat b) .\label
{eq:si-biasdriven}
\end{equation}
\item[(iii)] If, on the other hand, $T^{2}/(T^{2} + q^{2}\sefac
{T}{\mathrm{df}}^{2}) < R < 1$,
then \textup{\eqref{eq:si-norho}} is
sharp: its right-hand side represents the union of
\textup{\eqref{eq:climfmla}} as $(\pcor,\teeW)$ ranges over the sensitivity
zone $[-R^{1/2}, R^{1/2}]\times[-T, T]$.
\end{enumerate}
\label{keyprop}
\end{proposition}

Proposition \ref{keyprop} expresses solutions to the constrained
optimization problems
% of combining the bias and variance expressions
% in Propositions \ref{prop:univarOVbias} and \ref{prop:univarOVSE} to
of determining the smallest $\hat\beta- q\SE(\hat\beta)$ and largest
$\hat\beta+ q\SE(\hat\beta)$ consistent with assumed restrictions on
$\pcor$ and $\teeW$: in part (i), the restrictions pertain only to~$\teeW$,
while parts (ii) and (iii) impose restrictions on $\pcor$
also. The proof of the proposition appears in the \hyperref[app]{Appendix}.

\begin{remarks*} (a) In many problems $\mathrm{df}$ will be large in comparison with
plausible values of $\teeW^{2}$, confining $\sefac{\teeW}{\mathrm{df}}$ and
$\sefac{T}{\mathrm{df}}$ to the immediate vicinity of 1. (b)~Part (ii) says that
if the
magnitude of $\pcor$ is assumed to be small or moderate, then the
extremes of the sensitivity interval correspond to speculation
parameter values sitting at extremes of the sensitivity zone---the
same extremes at which (signed) omitted variable bias is minimized or
maximized. According to part (iii), however, if $\pcor$ is permitted
to be large, then restricting attention to sensitivity parameter
values that maximize or minimize omitted variable bias may lead
the statistician to underestimate the proper extent of the sensitivity
interval.
\end{remarks*}

% The proof of Proposition \ref{keyprop}, given in the Appendix, also
% establishes that for $T^{2}/(T^{2} + q^{2}\sefac{T}{df}^{2}) < R < 1$
% the interval at right of \eqref{eq:si-biasdriven} is strictly
% contained in the full sensitivity interval corresponding to
% the sensitivity zone $[-R^{1/2}, R^{1/2}]\times[-T, T]$.

%s3.1 ###
\subsection{\texorpdfstring{Pegging the boundaries of the sensitivity zone.}{Pegging
the boundaries of the sensitivity zone}}
Because our analysis began with covariate selection, there are a number
of deliberately omitted variables that can be used to peg at least one
boundary of the sensitivity zone. As one might expect, the covariates
eliminated by the stepwise procedure add little to those variables
that were included in terms of prediction of the response, and it
is too optimistic to suppose of a genuinely unmeasured confounder
that its contribution to the outcome regression would be no greater
than that of measured covariates that variable selection would put
aside. On the other hand, confounding with $Z$ plays little role in
common stepwise procedures like the one we used, and the deliberately
omitted variables can be used to guide intuitions about plausible
values of~$\teeW$. We selected six of these whose partial
associations with RHC spanned the full range of such associations
among stepwise-eliminated covariates and used their $\teeW$-values to
delimit the first dimension of several sensitivity zones.
Table \ref{omittedvars} delineates the $\teeW$-part of the sensitivity
zone accordingly, choosing the bound $T$ on treatment confounding to
coincide with the magnitude of confounding, conditional on
stepwise-selected covariates, between the treatment and each of our
six covariates.

The benchmarking method, using known variables to determine
plausible values of $\teeW$, informs targeted speculation about the
potential effects of omitted variable bias. Using existing information
in this way, we can speculate about the effects of omitted covariates that
%are refinements of or
are of a similar nature to measured covariates---$\teeW$ benchmarks
extracted from partial demographic information might reasonably
predict the $\teeW$ values that would attach to additional demographic
variables, were they available. To calibrate intuitions about omitted
variables that are different in kind from included ones, reference
values for $\teeW$ might also be obtained from external data sets.

% \begin{figure}
% \begin{center}
% \includegraphics[height=14cm, width=12cm]{logCIplot}
% \caption{95\% confidence and sensitivity intervals for the effect of
%RHC on hospital stay for omitted variables comparable in confounding
%with RHC to 6 selected stepwise-eliminated variables. They are defined
%from the bottom up as follows: (i) the ``omitted'' variable is
%included into the regression, (ii) sensitivity intervals with $\pcor=1
%unrestricted sensitivity intervals. \label{CIplot} }
% \end{center}
% \end{figure}

%Table as of 2/1/09
%

%t3 ###
\begin{table}
\tabcolsep=0pt
\caption{95\% sensitivity intervals for the treatment coefficient, with the
putative unobserved variable's treatment confounding ($|\teeW|$)
hypothesized to be no greater than the treatment confounding of 6
deliberately omitted variables. The decrease it would bring to the
variance of response residuals is hypothesized to be no greater than
either of 2 index values, 1\% and 10\%, or is not restricted}\label{omittedvars}
\begin{tabular*}{\textwidth}{@{\extracolsep{\fill}}l@{\qquad}cc@{\qquad}ccc@{}}
\hline
&\multicolumn{2}{c}{\textbf{}}&\multicolumn{3}{c@{}}{\textbf{\% decrease in unexplained variation}}\\
&\multicolumn{2}{c}{\textbf{}}&\multicolumn{3}{c@{}}{$\bolds{(100\rho^2_{y\cdot
w|z\mathbf{x}})}$}\\[-6pt]
& & & \multicolumn{3}{c@{}}{\hrulefill}\\
\multicolumn{1}{@{}l}{\textbf{Variable}}& \multicolumn{2}{l}{\multirow{3}{50pt}[21pt]{\centering \textbf{Treatment confounding benchmark}}} & \multicolumn{1}{c}{\textbf{1\%}} &
\multicolumn{1}{c}{\textbf{10\%}}& \multicolumn{1}{c@{}}{\textbf{Unrestricted}} \\

\hline
Insurance class & 12.2 &most& (0.03, 0.20) & ($-$0.04, 0.26) & ($-$0.21, 0.43) \\
Respiratory eval. & \phantom{0}8.9 &some& (0.04, 0.19) & ($-$0.01, 0.23) & ($-$0.12, 0.35) \\
Mean blood press. & \phantom{0}8.6 &some& (0.04, 0.19) & ($-$0.01, 0.23) & ($-$0.12, 0.34) \\
Cardiovascular eval. & \phantom{0}8.5 &some& (0.04, 0.19) & ($-$0.01, 0.23) &($-$0.11, 0.34) \\
Weight (kg) & \phantom{0}6.1 &some& (0.04, 0.18) & \phantom{0.}(0.01, 0.21) & ($-$0.05, 0.28)
\\
Immunosuppression & \phantom{0}0.4 &least& (0.06, 0.16) & \phantom{0.}(0.06, 0.16) & \phantom{0.}(0.06, 0.16) \\
\hline
\end{tabular*}
\end{table}

Many analysts will have sharper intuition for potential covariates'
effect on~$R^2$, making it relatively easy to set plausible limits on
$\pcor^{2}$. Table \ref{omittedvars} considers $\pcor^2 \leq0.01$ or
$0.10$, which may be useful as general starting points. In the present
case study, for instance, when included covariates are removed and put
in the role of $W$, $\pcor^2=0.01$ corresponds approximately to the
second most predictive of them, and the strongest single predictor
(DNR order) gives $\pcor^2=0.03$. It appears that $\pcor^2 \leq0.1$ is
a rather conservative bound: it is difficult even to find sets of
included covariates that jointly contribute so much to the outcome
regression as this. Only by simultaneously placing all of the
covariates into the role of $W$, leaving the intercept alone in the
role of $X$, does $\pcor^2$ reach 0.10. Restoring these variables into
the regression, and barring scenarios of still-omitted variables for
which $\pcor^2$ exceeds 0.1, Table \ref{omittedvars} shows the result
of a positive treatment effect to be sensitive to omitted confounding
on par with some of the strongest of the included confounders, but
insensitive to confounding weaker than that. If benchmarking leads to
accurate guesses about the values of treatment confounding and
reduction in unexplained variance, and if the linear model would hold
were the omitted confounder added to the regressors, then 95\%
sensitivity intervals will have 95\% coverage, despite the variable
omission.

%s3.2 ###
\subsection{\texorpdfstring{Basis for sensitivity formulas.}{Basis for sensitivity
formulas}}
\label{sec:theor-results-underl}
Propositions \ref{prop:univarOVbias}, \ref{prop:univarOVSE} and
\ref{keyprop} extend better-known descriptions of bias in regression
coefficients' point estimates due to variable omission
[e.g., \citet{seber1977}, page 66] to interval estimates. They also
have antecedents in earlier literature on numerical adjustment of
multiple regression results for the addition or removal of a covariate
[\citet{cochran1938ooa}]. Of the three,
Proposition \ref{prop:univarOVbias}'s proof is the most illuminating.
It also conveys the flavor of the others, which appear in the \hyperref[app]{Appendix}.

Consider $\mathbf{X}$ to be a matrix containing a column of 1's (or
columns from which a column of 1's can be recovered as a linear
combination) and let $Y$, $Z$ and $W$ be column vectors of common
extent, equal to the number of rows of $\mathbf{X}$. An inner product
is defined as $(A,B):=\sum{w_i}{a_i}{b_i}/\sum{w_i}$, where $w_i$ is a
quadratic weight for the $i$th observation (in the case of unweighted
least squares regression, $w_{i} \equiv1$). Write $\mathbf{1}$ for the
column vector of 1s. For vectors $A$, $B$ and $C$, let $\blp
(A|B,C)$ represent the projection of $A$ into the subspace spanned by
$B$ and $C$. Variances and covariances are defined as follows:
${\sigma_{ab\cdot c}} := (A-\blp(A| C), B-\blp(B| C))$,
$\sigma_{a\cdot c}^{2} = \sigma_{aa\cdot c}$; $\sigma_{ab}=
\sigma_{ab\cdot\mathbf{1}}$, $\sigma_{a}^{2}= \sigma_{a\cdot
\mathbf{1}}^{2}$. Partial correlations are then given as follows:
$\rho_{ab} := {\sigma_{ab}}/(\sigma_{a}\sigma_{b})$; $\rho_{ab\cdot c}
:= {\sigma_{ab\cdot c}}/(\sigma_{a\cdot c}\sigma_{b\cdot c})$. Denote
the degrees of freedom available for estimating $b$ as $\mathrm{df} = n-m-1$,
where $m=\operatorname{column.rank}(X)$. The nominal standard error
estimates for $\hat{b}$ and $\hat{\beta}$ [cf.
(\ref{eq:studyresult}) and (\ref{eq:desiredresult})] are then
%
%e8 ###
\begin{equation}\label{eq:stderrordef}
\SE(\hat{b}) = \mathrm{df}^{-1/2}\frac{\sigma_{y\cdot z\mathbf{x}}}{\sigma_{z\cdot
\mathbf{x}}} \quad  \mbox{and} \quad  \SE(\hat{\beta}) =
(\mathrm{df}-1)^{-1/2}\frac{\sigma_{y\cdot z\mathbf{x}w}}{\sigma_{z\cdot\mathbf{x}w}} .
\end{equation}

\begin{pf*}{Proof of Proposition \protect\ref{prop:univarOVbias}}
To show $\hat b - \hat\beta= \SE(\hat b)\teeW\pcor$, write
%
%e9 ###
\begin{equation}\label{eq:Wwithobs}
\blp(W| Z, {\mathbf{X}}) =: B^*Z+C^*{\mathbf{X}}^t .
\end{equation}

Using \eqref{eq:Wwithobs} to project the OLS estimate of regression
(\ref{eq:desiredresult}) onto the span of $(\mathbf{X}, Z)$ and then
comparing to (\ref{eq:studyresult}) gives $\hat{b}- \hat{\beta} =
B^*\hat{\delta}$, a well-known result [\citet{seber1977}, page 66].% We
%first express this bias in terms of partial correlations of $W$ with
%$Y$ and $Z$ given $\mathbf{X}$, or $\rho_{wy\cdot{\mathbf{x}}}$ and $\rho_{wz
%the product.

Write $W^{\perp_{\mathbf{x}}}$ for $W-\blp(W| \mathbf{X})$,\vspace*{1pt} $Z^{\perp_{\mathbf
{x}}}$ for $Z-\blp(Z| {\mathbf{X}})$, $Y^{\perp_{z{\mathbf{x}}}}$ for $Y-\blp
(Y| Z, {\mathbf{X}})$, and $W^{\perp_{z{\mathbf{x}}}}$ for $W-\blp(W| Z, {\mathbf
{X}})$. Then $\blp(W^{\perp_{\mathbf{x}}}|Z^{\perp_{\mathbf{x}}}) = B^*Z^{\perp
_{\mathbf{x}}}$, and
$\blp(Y^{\perp_{z{\mathbf{x}}}}| W^{\perp_{z{\mathbf{x}}}}) = \hat{\delta
}W^{\perp_{z{\mathbf{x}}}}$. These formulas imply $B^* = \sigma_{wz \cdot
\mathbf{x}}/\sigma^{2}_{z \cdot\mathbf{x}}$ and
$\hat{\delta} = {\sigma_{yw\cdot z{\mathbf{x}}}}/{\sigma^{2}_{w\cdot z{\mathbf
{x}}}}=\rho_{yw\cdot z{\mathbf{x}}} \sigma_{y\cdot z{\mathbf{x}}}/\sigma_{w
\cdot z{\mathbf{x}}}$, so that $\hat b - \hat\beta= B^{*}\hat{\delta}$
can be written as the product of ${\sigma_{y \cdot z{\mathbf{x}}}}/{{\sigma
}_{z \cdot{\mathbf{x}}}}$, $\sigma_{wz\cdot{\mathbf{x}}}/(\sigma_{z\cdot{\mathbf
{x}}}\sigma_{w\cdot z{\mathbf{x}}})$ and $\rho_{yw \cdot z{\mathbf{x}}}$.
Introducing mutually canceling factors of $(\mathrm{df})^{\pm1/2}$ to the first
and second of these and applying (\ref{eq:stderrordef}) turns this into
the product of $\SE(\hat b)$, $(\mathrm{df})^{1/2}\sigma_{wz\cdot{\mathbf
{x}}}/(\sigma_{z\cdot{\mathbf{x}}}\sigma_{w\cdot z{\mathbf{x}}}) $ and $\rho
_{yw \cdot z{\mathbf{x}}}$.\vspace*{1pt}
But $\teeW$ is just the ratio of $\sigma_{wz \cdot{\mathbf{x}}}/\sigma^{2}_{w \cdot{\mathbf{x}}}$ to $\sigma_{z \cdot w{\mathbf{x}}}/[(\mathrm{df})^{1/2}\sigma
_{w \cdot{\mathbf{x}}}]$, which simplifies to the second of these terms,
by way of the identity $\sigma^{2}_{z\cdot\mathbf{x}} \sigma^{2}_{w
\cdot z\mathbf{x}} = \sigma^{2}_{w \cdot\mathbf{x}} \sigma^{2}_{z
\cdot w \mathbf{x}} $ (an algebraic consequence of the definition of
$\sigma^{2}_{a \cdot c}$).
The result follows.
\end{pf*}

%s4 ###
\section{Extensions}
\label{sec:extens-applic}

As it is presented in Section \ref{sec:effect-estim-acco}, our method
explores sensitivities of covariance-adjusted estimates of a main
effect to the omission of a single covariate. It may appear to be
limited, then, to effect estimates made by linear covariate
adjustment, without interaction terms or other allowances for
heterogeneity of treatment effects, and to hidden bias scenarios
involving omission of a single variable, rather than several. Such an
appearance would be misleading.

%s4.1 ###
\subsection{\texorpdfstring{Several variables omitted at once.}{Several variables omitted at
once}}
\label{sec:severalOVs}

Suppose now that $W$ denotes not one but several omitted variables, or
that it represents a single nominal variable with~3 or more levels, so
that its encoding in terms of a design matrix would require 2 or more
columns, and 2 or more degrees of freedom. Results previously
presented still describe potential effects of $W$'s omission, if
$\teeW$ is reinterpreted in a natural way. (The sensitivity parameter
$\pcor^{2}$ retains its
original interpretation, as the proportionate decline in unexplained
variance from including $W$ as a regressor.)

When $Z$ is regressed on $X$ and a multivariate $W$, there is no one
$W$-coefficient and corresponding $t$-statistic. The natural analogue
of such a statistic is the ANOVA $F$-statistic comparing regression
fits with and without $W$, $\effW$; for univariate $W$, $\effW=
\teeW^{2}$, as is well known. When $\operatorname{rank}(W)>1$, define the
omitted variables' treatment confounding, again denoted $\teeW$, as
the positive square root of $[(k)(\mathrm{df})/(\mathrm{df}+1-k)]\effW$.
Proposition \ref{prop:univarOVbias} then gives the following.

\begin{corollary}\label{prop:mvOVbias}
Suppose $R^2_{y \cdot z\mathbf{x}}<1$, $t^{2}_W$ is finite, and $\operatorname
{rank}(W)=k>1$. Then $(\hat{b} - {\hat{\beta}})^2 \leq{\hat{V}}(\hat
{b}) [{(k) (\mathrm{df})}({\mathrm{df}+1-k})^{-1}]{F_W}\pcor^{2}$ or, equivalently,
\[
|\hat{b} - {\hat{\beta}}|\leq \SE(\hat b) \teeW|\pcor| .
\]
\end{corollary}

\begin{pf}
Without loss of generality, $W$ is uncorrelated with $Z$ and $X$: if
not, replacing $W$ with $W -\blp(W|X,Z)$ leaves $Z$-coefficients and
their standard
errors unchanged. Define $\tilde{W} = \blp(Y^{\perp{\mathbf{x}},
z}|W)$, where $Y^{\perp{\mathbf{x}}, z} = Y - \blp(Y
| X, Z) $. Again without loss of generality, $W = (\tilde{W}, W_2,
\ldots, W_k)$, where $\tilde{W} \perp(W_2,\ldots, W_k) $. Writing
%
%e10 ###
\begin{equation}\label{eq:mvdesiredresult}
\blp(Y| Z, {\mathbf{X}}, W) =: \hat{\alpha}+\hat{\beta}Z +
\hat{\gamma}{\mathbf{X}}^T + \hat{\delta}_1 \tilde{W} + \hat{\delta}_2 W_2
+ \cdots+ \hat{\delta}_k W_k   ,
\end{equation}
it is immediate that $\hat{\delta}_2, \ldots, \hat{\delta}_k=0$, since
$W_2, \ldots, W_k$ are orthogonal to\break $\blp(Y^{\perp{\mathbf{x}}, z} | W)$,
and hence orthogonal to $Y^{\perp{\mathbf{x}}, z}$. Projecting
(\ref{eq:mvdesiredresult}) onto the span of $(Z, \mathbf{X})$, and then
equating the $Z$-coefficient in what results with the $Z$-coefficient
in (\ref{eq:studyresult}) yields
%
%e11 ###
\begin{equation} \label{eq:1}
\hat\beta+ \hat{\delta}_1 B^*_1 = \hat b ,
\end{equation}
where $B^*_1$ is defined by $\blp(\tilde{W} |Z, X) = B^*_1 Z + C^* X$.
In other words, $\hat b$ and $\hat\beta$ are related just as they
would have been had $W$ been of rank 1, rather than $k$, consisting
only of $\tilde W$.

We record some entailments of the definitions of $\pcor$, $\teeW$ and
$\effW$ in a lemma, proved in the \hyperref[app]{Appendix}:
\begin{lemma}\label{multidimWlemma}
Suppose $R^2_{y \cdot z\mathbf{x}}<1$, $\teeW^2$ is finite, and $\operatorname
{rank}(W)=k$. Then:
\[
\mathrm{(1)}\quad \pcor^2 = \pcortilde^2;\qquad
\mathrm{(2)}\quad \teeWtilde^2 \leq k \frac{\mathrm{df}}{\mathrm{df}+1-k}\effW.
\]
\end{lemma}

The desired result now follows from (\ref{eq:1}), Proposition \ref
{prop:univarOVbias} and Lemma \ref{multidimWlemma}.
\end{pf}

When $\operatorname{rank}(W)>1$ we have the following variant of Proposition \ref
{prop:univarOVSE}, proved in the \hyperref[app]{Appendix}.

\begin{proposition}\label{prop:mvOVSE}
Suppose $R^2_{y \cdot z\mathbf{x}}<1$, $\teeW^2$ is finite, and $\operatorname
{rank}(W)=k$, $k>1$. Then
%
%e12 ###
\begin{equation} \label{eq:mvOVSE}
\hat{V}(\hat{\beta})=\hat{V}(\hat{b})\biggl[1+\frac{k + \teeW
^{2}}{\mathrm{df}-k}\biggr] (1- \pcor^2).
\end{equation}
\end{proposition}

Because the sensitivity intervals in Proposition \ref{keyprop} follow
algebraically from the bias and standard error representations
(\ref{eq:biasexpr}) and (\ref{eq:univarOVSE}),
they are valid for $W$ of arbitrary rank. The proofs of Proposition \ref
{keyprop} and the following are essentially the same.
\begin{proposition} \label{mvkeyprop}
Proposition \textup{\ref{keyprop}} continues to hold if
$\operatorname{rank}(W)=k>1$, provided
$\sefac{t}{d}$ is read as $[1+(k+t^{2})/(d-k) ]^{1/2}$
and $\teeW$ is read as $\{[k(\mathrm{df})/(\mathrm{df}+1-k)]\effW\}^{1/2}$.
\end{proposition}

Some of the hypothetical omissions discussed in
Section \ref{sec:effect-estim-acco} are of the type for which
Proposition \ref{mvkeyprop} is needed. The variable
``Insurance class'' appearing in Table~\ref{omittedvars}, for example,
is a nominal variable with 6 categories, consuming 5 degrees of
freedom when added as a regressor. Its treatment confounding,
$\teeW$, was calculated as the appropriately rescaled square-root of
the $F$-statistic
comparing the linear regression of $Z$ on $X$ and it to the regression
of $Z$ on $X$ alone, about $2.24$.

%s4.2 ###
\subsection{\texorpdfstring{Treatment effects differing by subgroup.}{Treatment
effects differing by subgroup}}
\label{sec:heterog-te}
Recall from Section \ref{ssec:OVbias} that of the 33 $X$-variables
selected as covariates for the regression of length of stay on RHC,
DNR status most reduced $R^2$. Patients with do-not-resuscitate orders
suffered 25\% greater mortality during the study than other patients,
probably making it inevitable that their outcomes on this variable
should systematically differ from patients without such orders. It is
natural to suspect that effects of RHC might differ for them as well.
In this case our linear models require interactions between RHC and
DNR status---and perhaps other interactions as well, for that
matter, but it suffices for us to restrict attention to the treatment
interaction with a single binary moderating variable, as all issues
pertaining to sensitivity analysis arise in this simplest case. \citet
{marcus1997} explores related problems.

Supplementing the regression of length of stay on covariates and RHC
with an interaction between RHC and DNR status gave quite revealing
results. The additional right-hand side variable, an indicator of RHC
and DNR simultaneously, bears a coefficient of $-0.43$ and a
$t$-statistic of $-5$: it appears that the model devoting a single
term to the treatment obscured significant effect heterogeneity.
Correspondingly, the main RHC coefficient, interpretable as the effect
for patients without DNR orders, is larger ($+0.15$) than it was in
earlier analyses without interactions ($+0.11$); its standard error
increases slightly, from $0.026$ to $0.027$.

To subject these results to sensitivity analysis, we again use
index values to limit the impact of the hypothesized omitted covariate
on $R^2$: $\pcor^2\leq0.10$ remains a generous limit, as
simultaneously adding all 33 covariates to the regression of length of
stay on RHC (now interacted with DNR status) decreases unexplained
variation by only slightly more, about 11\%. To set suitable limits
on the omitted variable's treatment confounding, imagine for the
moment that the inclusion of an interaction term had been handled
somewhat differently: rather than adding $ZX_1$, the product of RHC
and DNR indicators, add in $\tilde{X} = ZX_1 - \blp(ZX_1 | X, Z)$.
The coefficient of this variable lacks any straightforward
interpretation, but its addition to the right-hand side of the
equation has precisely the same effect on remaining coefficients as
would the addition of $ZX_1$ itself. To benchmark
treatment confounding, we would then regress $Z$ on $X$ and
$\tilde{X}$. By construction, however, $\tilde{X}$ is orthogonal to
$Z$ and $X$, so that it itself earns a $t$-statistic of $0$ in this
fitting, and its inclusion has no effect other than to remove a single
residual degree of freedom. In other words, the $\teeW$-benchmarks
extracted by regressing $Z$ on $X$ alone, used for sensitivity
analysis of the RHC effect in the absence of interactions, serve just
as well here after multiplication by the factor $[(\mathrm{df} +
1)/(\mathrm{df})]^{1/2}$ (which in this case is effectively 1). The first 2
columns of Table \ref{tab:SIwithInteractions} use benchmarks gathered
in this way, finding the conclusion that RHC increases lengths of time
in the
hospital for patients without DNR orders to be a bit less sensitive to
hidden bias than was the analogous conclusion for the analysis
assuming homogeneous treatment effects, in Table \ref{omittedvars}.

% latex table generated in R 2.8.0 by xtable 1.5-4 package
% Wed Feb 18 19:11:49 2009
%
%t4 ###
\begin{table}
\tabcolsep=0pt
\caption{Sensitivity intervals for subgroup effects and for weighted
average effects}\label{tab:SIwithInteractions}
\begin{tabular*}{\textwidth}{@{\extracolsep{4in minus 4in}}ld{2.1}cd{2.1}cd{2.1}c@{}}
\hline
&\multicolumn{6}{c@{}}{\textbf{Treatment-confounding benchmarks found with
estimated}}\\[-6pt]
&\multicolumn{6}{c@{}}{\hrulefill}\\
&\multicolumn{2}{c}{\textbf{Effect on patients}}&\multicolumn{2}{c}{\textbf{Effect on patients}}&\multicolumn{2}{c}{\textbf{ETT-weighted}}\\
&\multicolumn{2}{c}{\textbf{w/out DNR order}}&\multicolumn{2}{c}{\textbf{with DNR
order}}&\multicolumn{2}{c}{\textbf{average effect}}\\[-6pt]
&\multicolumn{2}{c}{\hrulefill}&\multicolumn{2}{c}{\hrulefill}&\multicolumn{2}{c@{}}{\hrulefill}\\
\textbf{Variable}& \multicolumn{1}{c}{$\bolds{\teeW}$} &\multicolumn{1}{c}{$\bolds{\rho_{y \cdot w|z\mathbf{x}}^2 \leq0.1}$}&
\multicolumn{1}{c}{$\bolds{\teeW}$} &\multicolumn{1}{c}{$\bolds{\pcor^2 \leq0.1}$}& \multicolumn{1}{c}{$\bolds{\teeW}$}&
\multicolumn{1}{c}{$\bolds{\pcor^2 \leq0.1}$}\\
\hline
Insurance class & 12.2 & ($-$0.01, 0.30) & 11.9 & ($-$0.74, 0.17)\phantom{0.} & 12.2& ($-$0.03, 0.27) \\
Respiratory eval. & 8.9 & \phantom{0.}(0.02, 0.27) & 8.1 & ($-$0.64, 0.08)\phantom{0.} & -4.1 & \phantom{0.}(0.03, 0.20) \\
Mean blood press. & -8.6 & \phantom{0.}(0.02, 0.27) & -8.2 & ($-$0.64, 0.08)\phantom{0.} &-5.1& \phantom{0.}(0.03, 0.21) \\
Cardiovascular eval. & -8.5 & \phantom{0.}(0.02, 0.27) & -7.7 & ($-$0.63, 0.07)\phantom{0.} &-7.1 & \phantom{0.}(0.01, 0.23) \\
Weight (kg) & 6.1 & \phantom{0.}(0.05, 0.25) & 6.4 & ($-$0.60, 0.03)\phantom{0.}& -5.3& \phantom{0.}(0.03, 0.21) \\
Immunosuppression & 0.4 & \phantom{0.}(0.09, 0.20) & 0.4 & ($-$0.44, $-$0.12) & -5.9 & \phantom{0.}(0.02, 0.22) \\
\hline
\end{tabular*}
\end{table}

When we instructed it to include the RHC--DNR interaction among the
explanatory variables, our software might equally well have added an
indicator of RHC and the absence of DNR, $Z(1-X_1)$. In this case,
the main effect would be interpretable as the effect of RHC for patients
\textit{with}, rather than without, DNR orders. It follows that had
that effect been the object of our interest, we could construct a
sensitivity analysis for it in the same manner as just above, by
persuading our regression program to expand the interaction
differently. In actuality, things are still simpler than that; we
really only need to take ordinary care in interpreting the regression
results, and somewhat modify the benchmarking equation. The effect
for patients with DNR orders is the sum of the main RHC effect and the
RHC-and-DNR effects, $0.148+ (-0.430)= -0.28$, with estimated variance
equal to the sum of the two estimated variances and twice their
covariance, $0.0065 = (0.080)^2$. In parallel, to benchmark treatment
confounding for this analysis, regress the sum of the main RHC
indicator and the interaction term, the product of RHC and DNR
indicators, on covariates. This gives somewhat different results
than did the benchmarking for the RHC effect on patients without DNR
orders, which omitted the interaction term from the left-hand side of
its regression equation; compare the middle and left columns of
Table \ref{tab:SIwithInteractions}.

The same approach yields a sensitivity analysis for any target
parameter representable as a linear combination of main effect and
interaction terms. Take the effect of treatment on the treated, or
ETT, parameter, considered from within a model permitting treatment
effects to vary within specified subgroups. If the groups are, for
simplicity, patients with and without DNR orders, then since~7\% of
patients receiving RHC had DNR orders, the ETT can be represented as
the main effect plus $0.07$ times the DNR--RHC interaction effect,
estimated as $0.148+ (0.07)-0.430=0.118$, with corresponding standard
error $0.026$. For benchmarking, regress on covariates the RHC
indicator plus 0.07 times the product of RHC and DNR indicators, with
results as given in the rightmost two columns of
Table \ref{tab:SIwithInteractions}.

%s4.3 ###
\subsection{\texorpdfstring{Propensity-adjusted estimates of treatment
effects.}{Propensity-adjusted estimates of treatment
effects}}
\label{sec:ppty-adjusted}
Regression adjusts between-group comparisons by attempting to remove
adjustment variables' contributions from the outcomes before comparing
them. In contrast, adjustments based on propensity scores attempt to
divide the sample into strata within which treatment and control
subjects have similar distributions of covariates. We estimated
propensity scores using all 50 of the SUPPORT data's covariates in a
logistic regression [\citet{rr1984}], finding six equally-sized
subclasses made treatment-control differences on the covariates jointly
insignificant at level
$\alpha=0.10$ [\citet{bh2008}]. One can couple such a stratification with linear
modeling to estimate treatment effects.
In the simplest variant, responses are regressed on the treatment and
fixed stratum effects. Fitting such a model to the SUPPORT data gives
an RHC effect similar to what was estimated after ordinary covariance
adjustment, as in Section \ref{sec:first-pass-regr}, but with somewhat
larger standard errors.

The main assumption for this model is that so far as the outcome,
length of stay, is concerned, the only systematic difference between
the RHC and non-RHC patient in a propensity stratum is RHC itself.
Relax this assumption in favor of another to the effect that so far as
differences between outcomes and their projections onto an omitted
variable are concerned, within strata RHC and non-RHC patients do not
systematically differ.
% more formally,
% $(Y_{t}, Y_{c}) \perp Z | S$, where $Y_{t}$ and $Y_{c}$ are potential
% responses to treatment and control and $S$ is an indicator of
% propensity stratum, is relaxed to $(Y_{t}- \blp(Y|W), Y_{c} -
% \blp(Y|W)) \perp Z | S$
Were the omitted variable to become available, we could adjust
by adding it to the explanatory side of the linear model.
Without it, we can do sensitivity analysis.

Benchmarking treatment-confounding levels takes a bit more
effort than before: rather than simply regressing $Z$ on covariates
and recording their $t$- or $F$-statistics, we have to account in some
way for the propensity stratification. We do this by removing the
covariates, one at a time, from the propensity model, after each
removal subclassifying the sample into sextiles, as before, but now
using the modified propensity score; then regressing $Z$ on the
withheld covariate and on the propensity strata in order to associate
a $t$- or $F$-statistic with that covariate. Results of this process
appear in Table \ref{proptab}.

The results exhibit a striking pattern: adjustment based on
propensity scores gives causal inferences that are far less sensitive
to omitted variables than does regression-based covariate adjustment.
With it, one can expect less residual confounding with the treatment
than with covariate adjustment, as seen in smaller~$\teeW$-values on
the propensity-score side of the table. In a sense, propensity scores
focus on confounding with the treatment, whereas covariate adjustment
focuses on covariates and the response. Recall from Section \ref
{sec:effect-estim-acco} that while both matter to omitted
variable bias, confounding with the treatment is both more difficult
to pin down and potentially more pernicious. It stands to
reason that while propensity adjustment may pay a slight
penalty up front, in terms of somewhat larger standard errors than
covariance adjustment, it offers a greater return downstream, in
reduced sensitivity to hidden bias.

%Table as of 2/11/09
%

%t5 ###
\begin{table}
\tabcolsep=0pt
\caption{
Sensitivity intervals for the treatment effect after ordinary
covariance and propensity-score adjustment, illustrating that
propensity adjustment better limits sensitivity to the omission of
adjustment variables. For covariance adjustment, $\teeW$ is limited by
the confoundedness with the treatment of 6 variables that had been
eliminated by a preliminary variable-selection procedure, as in
Table \textup{\protect\ref{omittedvars}}; for propensity adjustment, limits on treatment
confounding are set by separately removing each of these and
calculating their $\teeW$'s after propensity adjustment for remaining
variables}\label{proptab}
\begin{tabular*}{\textwidth}{@{\extracolsep{\fill}}ld{2.1}ccccc@{}}
\hline
&\multicolumn{3}{c}{\textbf{OLS regression}} &\multicolumn{3}{c@{}}{\textbf{Propensity adjusted
regression}}\\[-6pt]
&\multicolumn{3}{c}{\hrulefill} &\multicolumn{3}{c@{}}{\hrulefill}\\
& \multicolumn{1}{c}{{$\bolds{|t_W|}$}} & {$\bolds{\rho^2_{y \cdot w|z\mathbf{x}}\leq0.01}$}& {$\bolds{\rho^2_{y
\cdot w|z\mathbf{x}}\leq0.1}$}& {$\bolds{|t_W|}$} & {$\bolds{\rho^2_{y \cdot w|z\mathbf{x}}\leq
0.01}$}& {$\bolds{\rho^2_{y \cdot w|z\mathbf{x}}\leq0.1}$}\\
\hline
Insurance class & 12.2 & (0.03, 0.20)&($-$0.04, 0.26) & 8.6 & (0.02,
0.18) & ($-$0.03, 0.23) \\
Respiratory eval. & 8.9 & (0.04, 0.19)&($-$0.01, 0.23) & 3.1 & (0.04,
0.17)& \phantom{0.}(0.02, 0.19) \\
Mean blood press. & 8.6 & (0.04, 0.19) &($-$0.01, 0.23) & 6.8 & (0.03,
0.18)& ($-$0.01, 0.22) \\
Cardiovascular eval. &8.5 & (0.04, 0.19) &($-$0.01, 0.23) & 5.4 &
(0.03, 0.17)& (0, 0.20)\phantom{.} \\
Weight (kg) & 6.1 & (0.04, 0.18)&\phantom{0.}(0.01, 0.21) & 5.1 & (0.03, 0.18)&
\phantom{0.}(0.01, 0.21) \\
Immunosuppression & 0.4 & (0.06, 0.16)&\phantom{0.}(0.06, 0.16) &0.5 & (0.04,
0.16)& \phantom{0.}(0.04, 0.16) \\
\hline
\end{tabular*}
\end{table}
%s5 ###
\section{Summary}
\label{sec:summary}

For effect estimates adjusted for covariates using ordinary least
squares, impacts of covariate omission on point estimates and on
standard errors have been represented in terms of two statistics
relating the omitted variable to included ones, a measure of how
adding the variable would affect $R^2$ and a measure of its
association with the treatment variable given included variables. We
refer to the latter as the omitted variable's treatment-confounding
measurement. When generous limits on how the omitted variable would
affect $R^2$ can be defended, they yield far less pessimistic
assessments of sensitivity than would be possible without such a
limit. Unlike the sensitivity ``parameter'' pertaining to $R^2$,
plausible limits on the treatment-confounding parameter are unlikely
to emerge from intuition alone; on the other hand, it is
straightforward and informative to determine study-specific benchmarks
for it using available data.

The changes to the treatment coefficient's point and error estimates
that the addition of an omitted covariate would cause
have been represented as multiples of its standard error. So these
representations yield error appraisals accounting for certain hidden
biases in familiar terms, as a multiple of the SE.
% Indeed, they
% entailed a representation of the union of intervals
% that could have resulted had an omitted variable of given
% specifications been restored and a confidence interval for the focal
% variable then been constructed using the usual studentization
% technique.
The method adapts readily to scenarios of multivariate
omission, heterogeneous treatment effect and combinations of
regression with propensity scores.

\begin{appendix}\label{app}

%s6 ###
\section*{Appendix: Proofs}

\begin{pf*}{Proof of Lemma \protect\ref{multidimWlemma}}
Under the conditions of the lemma, (1) and (2) can be established: (1)
In a regression of $Z$ and ${\mathbf{X}}$ on $Y$, adding $\tilde{W}$ has
the same effect on the $Z$-coefficient and model $R^2$ as adding $W$.
Thus, (1) holds.

\noindent(2) Furthermore, $\tilde{W} \in\operatorname{span}(W)$, so $\tilde
{W}$ explains no more variation in $Z$ than does $W$. Thus, $R^2_{z
\cdot w|x} \geq R^2_{z \cdot{\tilde{w}}|x}$, which implies $\frac{\rho^2_{z \cdot w|x}}{1-\rho^2_{z \cdot w|x}} \geq\frac{\rho^2_{z
\cdot\tilde{w}|x}}{1-\rho^2_{z \cdot\tilde{w}|x}}$. By definition
of the ANOVA $F$-statistic, $\effW= [{(\sigma_{z\cdot\mathbf{x}}^{2} -
\sigma_{z\cdot\mathbf{x}w}^{2})/(k)}]/[{\sigma_{z\cdot\mathbf
{x}w}^{2}/(\mathrm{df}+1-k)}]$, or
%
%e13 ###
\begin{equation}
\effW= [({\mathrm{df} +1 - k })/{k}] [{\pcorzw^{2}}/({1- \pcorzw^{2}})]. \label{eq:3}
\end{equation}

% {\sigma_{z\cdot\mathbf{x}w}^{2}/(\mathrm{df}+1-k)}
% = \frac{\mathrm{df} +1 - k }{k} \frac{\pcorzw^{2}}{1- \pcorzw^{2}}
% \end{equation}

%In Section \ref{sec:severalOVs}, $\teeW$ is defined for multivariate
%$W$. Since $k = \mathrm{rank}(W)$, in case of univariate $W$, $
As $\operatorname{rank}(\tilde{W})=1$, $F_{\tilde{W}}=t_{\tilde{W}}^2 = \mathrm{df}
\frac{\rho^2_{z \cdot\tilde{w}|x}}{1- \rho^2_{z \cdot\tilde
{w}|x}}$. The result follows.
\end{pf*}

\begin{pf*}{Proof of Proposition \protect\ref{prop:mvOVSE}}
To relate $\SE(\hat{b})$ and $\SE(\hat{\beta})$, begin by rewriting
some of the variance terms:
$\sigma^2_{y \cdot z {\mathbf{x}}w} = {\operatorname{Var}}(Y^{\perp{z{\mathbf{x}}w}})
={\operatorname{Var}}(Y^{\perp{z{\mathbf{x}}}} - \blp(Y^{\perp{z{\mathbf
{x}}}}|W^{\perp{z{\mathbf{x}}}}))
= \sigma^2_{y \cdot z{\mathbf{x}}} - \sigma^2_{y \cdot z{\mathbf{x}}}\rho^2_{yw \cdot z{\mathbf{x}}}
= {\sigma}_{y \cdot z\mathbf{x}}^{2}({1 - \rho^2_{yw \cdot z{\mathbf{x}}}})$. Similarly, ${\sigma}_{z \cdot{\mathbf
{x}}w}^{2} = {\sigma}_{z \cdot{\mathbf{x}}}^{2}(1 - \pcorzw^{2}).$

Thus, by (\ref{eq:stderrordef}), $\SE(\hat{\beta})
=(\mathrm{df}-k)^{-1/2}({\sigma}_{y \cdot z {\mathbf{x}}}/{{\sigma}_{z \cdot{\mathbf
{x}}}})[({1 - \rho^2_{yw \cdot z{\mathbf{x}}}})/({1 - \rho^2_{zw \cdot
{\mathbf{x}}}})]^{1/2}$, or, invoking (\ref{eq:stderrordef}) again,
$\SE(\hat{b})[{\mathrm{df}}/(\mathrm{df}-k)]^{1/2}[({1 - \rho^2_{yw \cdot z{\mathbf
{x}}}})/({1 - \rho^2_{zw \cdot{\mathbf{x}}}})]^{1/2}$.

By \eqref{eq:3}, $({1 - \pcorzw^{2}})^{-1} = 1 + {k}({\mathrm{df}+1-k})^{-1}
\effW$.
Recall that $\teeW$ was defined for multivariate $W$ in Section \ref
{sec:severalOVs}, as a rescaling of $\effW^{1/2}$. Applying that
definition, $({1 - \pcorzw^{2}})^{-1} = ({\mathrm{df} + \teeW^{2}})({\mathrm{df}}^{-1})$.
The relationship \eqref{eq:mvOVSE} follows.
\end{pf*}

\begin{pf*}{Proof of Proposition \protect\ref{keyprop}}
Equation \eqref{eq:climfmla} comes directly from\vspace*{1pt} Propositions \ref
{prop:univarOVbias} and \ref{prop:univarOVSE}. For \eqref{eq:si-norho},
write $\hat{\beta} - q\SE(\hat{\beta}) = \hat{b} + l_{\teeW}(\arcsin
\pcor)\SE(\hat{b})$ and $\hat{\beta} + q\SE(\hat{\beta}) = \hat{b} +
u_{\teeW}(\arcsin\pcor)\SE(\hat{b})$, where $\arcsin\pcor\in\break(-\pi
/2, \pi/2)$ and $l_{t}(\theta) := -t \sin\theta- q \sefac{t}{\mathrm{df}}\cos
\theta$ and $u_{t}(\theta) := -t \sin\theta+\break q\sefac{t}{\mathrm{df}}\cos\theta
$. To maximize $u_{t}(\arcsin\rho)$ as $t$ ranges over $[-T, T]$ and
$\rho$ ranges over $[-1,1]$, maximize $t \mapsto\sup\{ u_{t}(\arcsin
\rho)\dvtx -1 \leq\rho\leq1\}$. We need only consider $t<0$. For such a
$t$, calculus shows that $u_{t}(\cdot)$ is concave unimodal on $(-\pi
/2, \pi/2)$, attaining its maximum at $\arctan\{-t/[q \sefac{t}{\mathrm{df}}]\}
$;\vspace*{1pt} with some algebra and trigonometry, $\sup\{ u_{t}(\arcsin\rho)\dvtx -1
\leq\rho\leq1\}$ is seen to be\break $({t^2+q^2\sefac{t}{\mathrm{df}}^{2}})^{1/2}$.
Consequently, $\sup\{ u_{t}(\arcsin\rho)\dvtx |t|\leq T, |\rho| \leq1\}
=\break
({T^2+q^2\sefac{T}{\mathrm{df}}^{2}})^{1/2}$. Similarly, as~$t$ and $\theta$
range over $[-T, T]$ and\break $(-\pi/2, \pi/2)$, $l_{t}(\theta)$ takes its
minimum value, $-({T^2+q^2\sefac{T}{\mathrm{df}}^{2}})^{1/2}$, at\break $(T, \arctan\{
T/[q \sefac{T}{\mathrm{df}}]\})$. Part (i) follows.

Restating slightly an intermediate conclusion, the maximizer of $\rho
\mapsto\break u_{t}(\arcsin\rho) $ over the domain $[-1,1]$ is $-t/[t^{2} +
q^{2}\sefac{t}{\mathrm{df}}^{2}]^{1/2}$. Under the condition of (iii), for each
$t \in[-T,T]$ this falls within the narrower domain $[-R^{1/2},
R^{1/2}]$. (iii) follows.

In (ii), $\pcor^{2} \leq R $. To maximize $u_{t}(\arcsin\rho)$ over a
domain that is symmetric in $t$, we again need only consider negative
$t$. For $t$ small enough in magnitude that $t^{2}/[t^{2} + q^{2}\sefac
{t}{\mathrm{df}}^{2}] \leq R$, the maximizer of $\rho\mapsto u_{t}(\arcsin\rho
) $ over the domain $[-1,1]$ falls inside the narrower domain
$[-R^{1/2}, R^{1/2}]$, and $\sup_{-R^{1/2} \leq\rho\leq R^{1/2}}
u_{t}(\arcsin\rho) = \sup_{-1 \leq\rho\leq1} u_{t}(\arcsin\rho) =
({t^2+q^2\sefac{t}{\mathrm{df}}^{2}})^{1/2} $.\break This function is increasing as a
function of $-t$. For $t$ such that $R \leq t^{2}/[t^{2} + q^{2}\sefac
{t}{\mathrm{df}}^{2}]$, because $u_{t}(\cdot)$ is concave unimodal with maximum
at a point,\break $\arctan\{-t/[q \sefac{t}{\mathrm{df}}]\}$, that falls outside of
$\{ \arcsin\rho\dvtx |\rho| \leq R^{1/2}\}$,\break $\sup\{ u_{t}(\arcsin\rho)\dvtx
{-R^{1/2} \leq\rho\leq R^{1/2}}\} = u_{t}(\arcsin R^{1/2}) =
-tR^{1/2} + q \sefac{t}{\mathrm{df}}(1 - R)^{1/2}$. This also is increasing as a
function of $-t$.\vspace*{1pt} For the unique $t<0$ such that $R = t^{2}/[t^{2} +
q^{2}\sefac{t}{\mathrm{df}}^{2}]$, $\sup\{ u_{t}(\arcsin\rho)\dvtx {-R^{1/2} \leq
\rho\leq R^{1/2}} \}$ is given by either of the two functions of $-t$,
which shows that they coincide at that point. In consequence, $\sup\{
u_{t}(\arcsin\rho)\dvtx {-R^{1/2} \leq\rho\leq R^{1/2}}\}$ is increasing
as a function of $-t$ for all $t <0$, so that the maximum of
$u_{t}(\arcsin\rho)$ for $|t| \leq T$ and $\rho^{2} \leq R$ is $\sup\{
-tR^{1/2} + q \sefac{t}{\mathrm{df}}(1 - R)^{1/2}\dvtx |t| \leq T\} = TR^{1/2} + q
\sefac{T}{\mathrm{df}}(1-R)^{1/2}$, as required for (ii). [Similar steps yield
the minimum of $l_{t}(\arcsin\rho)$.]
\end{pf*}
% Now suppose $\pcor^2 \leq g(t_w,q,c)$. For some $\rho$ such that
% $$0 \leq\rho\leq\sin \arctan({|\teeW|}[{{(\teeW^2 +
%otherwise it is $-\rho$ if $\teeW>0$. Similarly, $\operatorname{arg} \min_{
%if $\teeW>0$.

%left\{ \begin{array}{ll}
% \rho, & \teeW<0 \\
% -\rho, & \teeW>0 \\
% \end{array} .

% -\rho, & \teeW<0 \\
% \rho, & \teeW>0 \\
% \end{array} .

% To verify the relationship for $u$, note $u(\arcsin(\cdot))$ takes
%only one maximum as its argument ranges over $[-1, 1]$ at $\sin
%$u(\arcsin\pcor)$ must take its maximum as $\pcor$ ranges over $[-
%for $l$ is established by similar reasoning.
\end{appendix}

\section*{Acknowledgments}
The authors wish to thank: Derek Briggs, Stephen Fienberg, John
Gargani, Tom
Love, Tanya Moore, Paul Rosenbaum, Nanny Wermuth and an anonymous
reviewer, for helpful comments; and Connors et al., in particular,
F. Harrell, for making their data publicly available.

\begin{supplement}[id=supp]
\exhyphenpenalty-10000
\stitle{Code for computations discussed in the article\\}
\slink[doi]{10.1214/09-AOAS315SUPP}
\slink[url]{http://lib.stat.cmu.edu/aoas/315/supplement.zip}
\sdatatype{.zip}
\sdescription{Zip archive containing our documented R code and
instructions for obtaining Connors et al.'s data, in the form of a Sweave
and a corresponding PDF file.}
\end{supplement}

\printaddresses

\end{document}